\documentclass[12pt,a4paper]{article}
\usepackage[latin1]{inputenc}
\usepackage{amsmath}
\usepackage{amsfonts}
\usepackage{amssymb}
\usepackage{epsfig}
\begin{document}

%%%%%%%%%%%%%%%%%%%%%%%%%%%%%%%%%%%%%%%%%%%%%%%%%%%%%%%%%%%%%%%%%%%%%%%%%%%%%
\title{\textbf{\large {Effects of Plasma Drag on Low Earth Orbiting Satellites due to Heating of Earth's Atmosphere by Coronal Mass Ejections}}}

\author{{\normalsize{\bf Victor U. J. Nwankwo$^{1}$\footnote{Corresponding author: victornwankwo@yahoo.com}, and Sandip Kumar Chakrabarti$^{1,2}$}}\\
\vspace{0.1cm}
{\normalsize{victornwankwo@yahoo.com; chakraba@bose.res.in}}\\
{\normalsize{$^{1}$ S. N. Bose National Centre For Basic Sciences, Kolkata 700098, India}}\\
{\normalsize{$^{2}$ Indian Centre for Space Physics, Kolkata 700084, India}}}
\date{}

\maketitle

\begin{abstract}
Solar events, such as coronal mass ejections (CMEs) and solar flares,
heat up the upper atmosphere and near-Earth space environment. 
Due to this heating and expansion of the outer atmosphere by 
the energetic ultraviolet, X-ray and particles expelled from the sun,
the low  Earth-Orbiting satellites (LEOS) become vulnerable to
an enhanced drag force by the ions and molecules of the expanded atmosphere. 
Out of various types of perturbations, Earth directed CMEs play 
the most significant role. They are more frequent and intense 
during the active (solar maximum) phase of the sun's approximately 
11-year cycle. As we are approaching another solar maximum later 
in 2013, it may be instructive to analyse the effects of the past 
solar cycles on the orbiting satellites using the archival data 
of space environment parameters as indicators. In this paper, we 
compute the plasma drag on a model LEOS due to the atmospheric heating 
by CMEs and other solar events as a function of the solar parameters. 
Using the current forecast on the time and strength of the next solar 
maximum, we predict how an existing satellite orbit may be affected in the 
forthcoming years.\\

\end{abstract}

%%%%%%%%%%%%%%%%%%%%%%%%%%%%%%%%%%%%%%%%%%%%%%%%%%%%%%%%%%%%%%%%%%%%%%%%%%%%%
%% Main text
\section{Introduction}

Solar activity is long known to have a significant influence on the upper atmosphere and 
the near-Earth space environment which affects modern equipments and consequently human activities on Earth.
Solar activity has an approximate 11-year cycle, which includes a stage each of solar maximum and minimum. Normally, the sun emits a continuous stream of energetic particles and electromagnetic radiation 
which has various time scales and intensities. The major processes of the solar energetic emissions
are the solar flares and the coronal mass ejections (CMEs). Both of these processes are sporadic,
and are the primary causes of adverse space weather. These processes are more frequent and 
intense during the solar maximum but less in magnitude and frequency during solar minimum. 
It has been predicted by NASA \cite{NOAA09} that the current, i.e., 24th solar cycle will attain its peak later in 2013, typically lasting for about six months. Given that a severe space weather has a direct and generally adverse effects on low earth orbiting satellites (LEOS) on which many human activities depend, it is pertinent to compute the severity of the effects ahead of the events.\\

The space weather events are natural sources of hazards \cite{IRG10,Lld10}. 
Effects of the space weather on the Earth and space systems include satellite drag \cite{Gop09};
disruption and damage of modern electric power grids,
corrosion of oil/gas pipelines due to geomagnetic induced current (GIC),
satellite sensor degradation; radiation threat to crew of high-flying aircraft 
and astronauts \cite{IRG10,Gop09,NRC08}, degradation of precision of Global Positioning System's (GPS) measurement; operational anomalies in satellites, damage of critical electronics and degradation of solar arrays
due to exposure to energetic particles during solar particle events (SPE)
\cite{NRC08,Jea11} etc. Secondary effects arise due to inter-dependency of systems on near-Earth 
space missions. Societal and economic impacts are also part of the resultant risks \cite{IRG10,Lld10}.\\

A number of factors determine the effects of the space weather on the satellite orbits.
These include the phase of the 11-year solar cycle, the nature of the spacecraft orbit, 
the local time and the position of the satellite relative to Earth-Sun direction \cite{NOAA06}. 
Although impact probabilities (affecting the orbit) increase with the rising phase of the peak,
certain effects occur at all the phases (and/or stages) of the cycle. For instance, the Van Allen Radiation Belt 
(VARB) is a potential threat zone for satellites. The effect is more or less
independent of the phase of the cycle. Energetic particles, such as electrons 
are present during solar minimum as well and also affects the satellite system. 
It was reported \cite{NOAA06,Ioa01} that two Canadian satellites (including Anik E1) 
experienced debilitating upsets in 1994 at the end of a long duration of 
adverse condition of the space weather. Similarly, Telsat 410 
(AT \& T communication satellite) and Galaxy 4 also failed in 1997 and 1998 respectively \cite{Ioa01}. 
A notable space weather event that produced significant impact during past solar peak is the Carrington event of September 1859. 
This is perhaps the largest recorded geomagnetic storm \cite{IRG10,NRC08,Lld10,POST10}. 
This event significantly disrupted telegraph systems around the world for as long 
as eight hours. Two other similar events, albeit of lesser magnitude, occurred in March 1989 
and in October/November 2003 respectively \cite{IRG10,NRC08,Lld10,Ioa01,POST10}. 
The former, a geomagnetic storm that resulted to the collapse (in about ninety minutes) 
of the Hydro-Quebec (Canada) power system, and led to complete blackout of utility grid in North America for about nine hours.
The latter event caused long-hour power outage in Sweden, and in UK, temporarily changed the `compass north' 
by five degrees for six minutes \cite{POST10}. More than 30 satellite anomalies 
were reported as a consequence of this event, with one of Japan lost completely \cite{Lld10,POST10}.\\

Because of these plethora of evidences, there are justifiable concerns on the probable 
impact of the space weather due to the forthcoming solar maximum. In this section, we 
concentrate on studying the effects of the solar events systematically. In the next Section,
we briefly describe major solar events which are taken care of in our study. In \S 3, we
present the method of our analysis. In \S 4, we present the computation of the average orbital 
decay due to the solar events in question. In \S 5, we present our results and discussions. Finally,
we make our concluding remarks.

\section{Factors affecting Space Weather in solar system}

The Sun is clearly the primary cause of space weather condition in the solar system. 
In a solar wind, a stream of energized charged particles (primarily protons 
and electrons) constantly flows outward from the solar corona and hits the magnetic corona.
A momentary and sudden release of the magnetic energy, also known as the solar flares,  
and large-scale, high-mass, eruptions of plasma, widely known as the Coronal Mass Ejections,
are also created in the solar corona and are injected into the interplanetary space, which 
occasionally hits the Earth, especially when they are earth-pointing.  
Electromagnetic and particle flux radiation emitted during these processes cause
adverse conditions in space weather. Among all the events, CMEs are the most damaging, 
because they may scoop out up to $10^9$ tons of magnetized plasma \cite{Gop09} from the Sun and inject it into 
the interplanetary space, thereby increasing the chance of its interaction with the Earth.
While the solar flares inject energetic particles and radiations into the interplanetary space, 
CMEs propagate inside the solar wind and drive shock waves, which in turn accelerate energetic particles 
\cite{Gop09,NOGS06}. CMEs are sometimes associated with solar flares and prominence eruptions but do also occur without any of the processes. 
These ejecta interact with the Earth's magnetosphere perturbing it and often causing rapid changes in the Earth's magnetic field. 
This process result in  geomagnetic storms. A CME can reach the earth in about a day or more
depending on its speed of propagation \cite{Gop09}. The mechanism of acceleration and propagation of CMEs 
are not well understood \cite{Ruf05,Gea09,MnG05}, though efforts are on to 
understand them using various models \cite{Gea09}. 

\section{The Procedure of our analysis}

In this work, we study the time variation of different solar parameters as indicators of the Solar activity. 
We procured archival data for past space weather events and analysed them. We compute the plasma drag 
on a model satellite in lower earth orbit (LEO) during the events and predict using the solar cycle forecast, 
as to how the satellite orbit could be affected around the peak of next solar maximum. 
In addition to the CME catalog \cite{NASA01}, three solar parameters were used as tracers of the phase of a solar cycle,  
They are: a) observed daily solar flux and geomagnetic Ap indices \cite{NOAA02}, b) predicted monthly mean 
solar flux indices \cite{NOAA03}, and c) sunspot number \cite{NOAA02}. Our analysis covers approximately 
two cycles (1995-2019), including  a seven-year predicted quantities (2013-2019).
We made two-stage analysis and grouped the data according to their anticipated impact level: 
(i) 1995-2009, assumed to be the period between a solar minimum to the next (min-to-min), and 
(ii) 1999-2013, assumed to be the period between a solar maximum to the next (max-to-max). 
Further, we analyzed the data around the last solar maximum and minimum.  
The grouping is shown in Table \ref{tbl1} (also see, \cite{Pea04,Pol02} below). 
Our choice of fifteen years interval (between two minima and maxima) is not only 
for convenience, but also because (i) the solar cycle is not strictly of 11-year duration,
(ii) occurrence of a seemingly `double-peak' during some solar maximum stage (as could be observed in figure 1b); increases the interval between one peak to the next peak, and 
(iii) the influence of the `extended' sporadic solar activity after some solar maximum, 
such as the event of late 2003.\\

\begin{table}[ht]
 \begin{center}
     \small\addtolength{\tabcolsep}{-4.5pt}
\caption{\label{tbl1} Classified/grouped data according to anticipated impact level}
\footnotesize\rm
\begin{tabular}{|c|c|c|c|c|c|}%{\textwidth}{@{}l*{15}{@{\extracolsep{0pt plus12pt}}l}}
\hline
& \multicolumn{2}{|c|}{Solar flux (F10.7) index} & & \multicolumn{2}{|c|}{Geomagnetic Ap index}\\
\hline
 &Value & Classification & Group& Value & Classification\\
\hline
1 & 65-99 & Average solar activity (ASA) & A & 0-14 &Quiet/Avg mag. condition (QMC)\\
2 & 100-150 & Moderate solar activity (MSA) & B & 15-29 &Active magnetic condition (AMC)\\
3 & 151-200 & High solar activity (HSA) & C & 30-49 &Minor storm condition (MnSC)\\
4 & 201-250 & Very high solar activity (VSA) & D & 50-99 &Major storm condition (MjSC)\\
5 & 251-300 & Extreme solar activity (ESA) & E & $\geq 100$ &Severe storm condition (SSC)\\
\hline
\end{tabular}
 \end{center}
\end{table}

Solar flux index (F10.7), monitored at the $10.7$ cm wavelength is treated as a measure of the 
contribution from interactions and subsequent heating of the upper atmosphere by solar 
energetic particles and ultra-violet (UV) radiation during solar events. The geomagnetic 
planetary K index (from which the planetary A index is derived) represents the measure 
of the contribution from the additional atmospheric heating that happens during geomagnetic 
storms \cite{NOAA06,IRS99,Pea04}. Upper atmospheric expansion is a direct consequence of 
the heating as measured by these solar parameters. This causes the atmospheric density at higher altitude 
to increase resulting to an increase in drag on satellites especially those at LEO. The satellite orbit 
decays and causes its re-entry into the Earth, unless appropriate corrective measures are taken to 
stabilize its orbital parameters.  

\section{Computation of orbital decay due to plasma drag}

In order to compute the orbital decay of a satellite orbit, we apply a simple atmospheric 
model equation to begin with, and study the effects of the space environmental parameters (SEP) 
on it. For concreteness and without any loss of generality, we assumed the satellite
with an exposed surface area of a unit square meter ($1$ m$^2$) in all directions,  
to possess a mass of $100$kg and orbiting the Earth at an initial injected circular orbit of 
radius $400$km. We chose a spherical polar co-ordinate system ($r, \phi$) having origin $r=0$ at the 
center of the Earth and assume that the satellite always remained in the same plane (i.e., $\theta=$ constant).
The effects of the drag force were computed from three basic sets of equations. The first set 
consists of four coupled differential equations.

%\section{An equation}
\begin{eqnarray}
\dot{v_r} &=& - \frac{G M_e}{r^2} + r \dot{\phi^2}, \quad \dot{r} = v_r ,
\end{eqnarray}

\begin{eqnarray}
\nonumber  \ddot{\phi} &=& - \frac{1}{2} r \rho \dot{\phi^2} \frac{A_s C_d}{m_s}, \quad \dot{\phi} = v_\phi/r .
\end{eqnarray}
Here, $v_r$ and $v_\phi$ are the radial and tangential velocity components respectively. 
$G$ is the gravitational constant, $M_e$ mass of the Earth, $r$ is the instantaneous radius of the orbit,
$\rho$ atmospheric density, $A_s$ is the omni-directional projected area of the satellite, 
$m_s$ is the mass of the satellite and $C_d$ the drag coefficient at an altitude of $r$. Note that the 
drag force has been applied only to the tangential direction, since the velocity is very high only in that direction.
The four differential equations are solved by the fourth order Runge-Kutta method to obtain instantaneous
positions and velocity components of the satellite in an orbit. To measure the decay of the orbital radius 
per orbit, we assume that the energy is constant per orbit.

Let $E_{total}$ be the total energy (kinetic and potential) in a orbit. After one revolution, 
the energy changes to $E'_{total}$ which is given by \cite{USA10},
\begin{eqnarray}
E'_{total} = E_{total} - W_{drag},
\end{eqnarray}
\begin{eqnarray}
\nonumber E_{total} &=& \frac{m_s r v^2 - 2 G M_e m_s}{2r}, \quad v = \sqrt{v_r^2 + v_\phi^2},\\
\nonumber W_{drag} &=& \frac{1}{2} \rho A_s C_d v^2 s, \quad s = 2 \pi r,
\end{eqnarray}
where, $W_{drag}$ is the work done by the drag force per revolution, i.e., after a traversal of $s$ distance.
The input parameter $\rho$ in both sets of equations above are to be supplied by the following equation \cite{IRS99},
\begin{eqnarray}
\rho = 6 \times 10^{-10} \exp -\frac{(h - 175)}{H},
\end{eqnarray}
where, $T$ (in Kelvin), $m$ and $H$ (in km) are the exospheric temperature 
(as a function of the solar flux and geomagnetic Ap index), 
effective atmospheric molecular mass, and variable scale height 
respectively. $h$ (km) is the satellite's altitude. These quantities are given by,
\begin{eqnarray*}
\nonumber T &=& 900 + 2.5(F10.7 - 65) + 1.5(A_p)\\ 
\nonumber m &=& 27 + 0.012(h-200);\ 180\geq h \geq 500km;\\
\nonumber H &=& T/m .
\end{eqnarray*}
%  \overfullrule 5pt
%  \mathindent\linewidth\relax
%  \advance\mathindent-259pt

\section{Statistics of Solar events which affect Satellite Orbits }

Figures 1(a-d) show the distribution of solar activity parameters during 
1995-2019 period, which include the extrapolated data of 2013-2019 based on present state 
of the solar activity. In Fig. 1a, we show the daily distribution of the solar flux 
(left Y-axis) and the geomagnetic Ap Index (right Y-axis) by brown points and 
blue points respectively. The average values are drawn by red and yellow curves 
respectively. From the year 2013 onwards (from day number 6580)
the data is smooth as they are extrapolated results. The peak in the upcoming solar maximum is clearly shallow
than the one of immediate past. The duration of the present cycle is also expected to be shorter.
In Fig. 1b, we replot sunspot number and solar flux index (data smoothed over one month) from 1991 to 2019, 
where the data since 2013 being the predicted value \cite{NOAA03}.
We then superposed the smoothed Geomagnetic Ap Index numbers (right Y-axis). 
In Fig. 1c, we plot the number of CMEs per day and in 
Fig. 1d, we plot the linear speed of CMEs (in km/s).\\

\vspace{1cm}
\begin{figure}
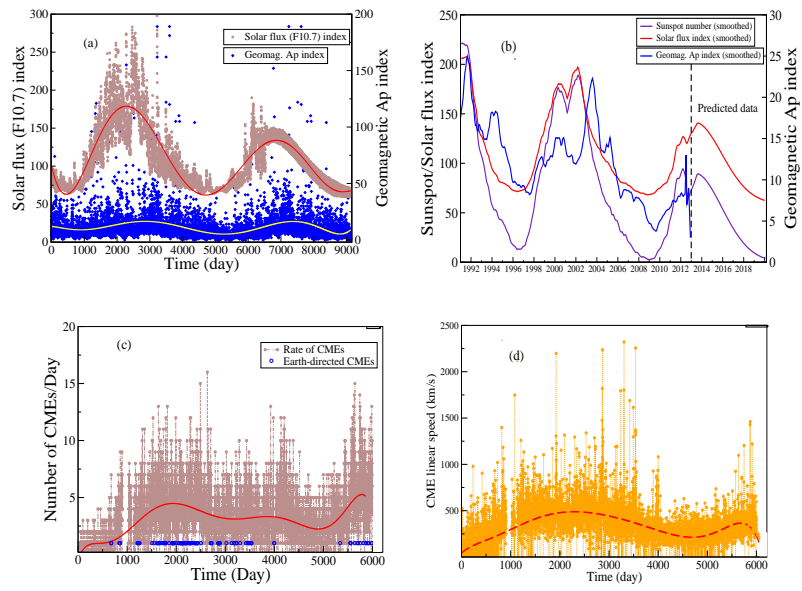

 \begin{center}
\includegraphics[height=3.5cm,width=5cm]{figure1a} \hspace{0.1cm}
\includegraphics[height=3.5cm,width=5cm]{figure1b}\\
\vspace{0.6cm}\includegraphics[height=3.5cm,width=4.5cm]{figure1c}
\hspace{0.4cm}\includegraphics[height=3.5cm,width=4.5cm]{figure1d}
 \end{center}
\caption{(a) Daily distribution of observed (1995-2012) and predicted (2013-2019) solar flux and geomagnetic Ap 
index (b) monthly mean distribution of observed (and predicted) sunspot number, 
solar flux and Ap index (c) Daily rate of CMEs and (d) CME mean linear speed during 1996-2012}
\label{fig1}
\end{figure}
 
The Figures clearly indicate that the magnetic condition is quieter (QAC) during solar minima but significantly increases during the solar maxima. Once `quiet', the magnetic activity increase gradually 
as the solar activity increases, but the mean peak of geomagnetic index does not have a direct 
correlation with the solar maximum. The fall in magnetic activity is equally gradual and the 
process last even after the solar maximum is over. It is clear that any of the magnetic conditions as classified in Table \ref{tbl1} (such as, QAC, AMC, MnSC, MjSC or SSC) in the magnetosphere may occur at any 
stage of the solar cycle but varies in frequency. Solar 
flux (F10.7) index of VSA and ESA class events rarely occur during 
the solar minimum (see also Fig. 4d below). Figure 1c showed that the rate of CMEs and the mean linear 
speed increases significantly around a solar peak, and therefore the probability of having 
more Earth-directed CMEs is also increased. A seemingly `double-peaked' solar maximum has 
been observed in the past sunspot data \cite{NASA07,SID08}. A similar signature is also observed 
in Figure 1b. Indeed, the actual data shows that 1661, 1500 and 1700 CMEs occurred in 2000, 
2001 and 2002 respectively - a clearly convincing example of a 'double-peak' in CME 
number variation. Observed sunspot number and solar flux index (smoothed over one month period)
have similar signatures during that period.\\

\begin{figure}
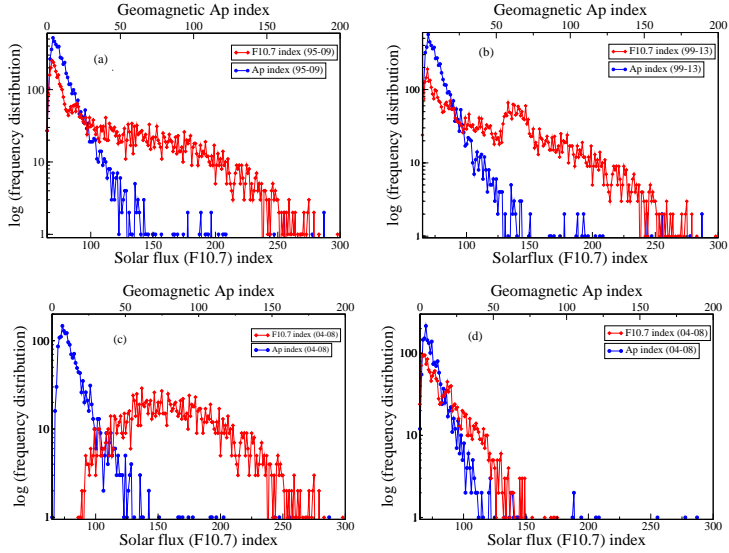

 \begin{center}
\includegraphics[height=3.5cm,width=4.5cm]{figure2a} \hspace{0.2cm}
\includegraphics[height=3.5cm,width=4.5cm]{figure2b} \\
\vspace{0.2cm}\includegraphics[height=3.5cm,width=4.5cm]{figure2c}
\hspace{0.2cm}\includegraphics[height=3.5cm,width=4.5cm]{figure2d}
 \end{center}
\caption{Frequency of occurrence of solar flux and geomagnetic Ap index (a) min-to-min: 
1995-2009 (b) max-to-max:1999-2013 (c) around the last solar maximum: 1999-2003 and (d) the last solar minimum: 2004-2008}
\label{fig2}
 \end{figure}

\begin{figure}
 \begin{center}
\includegraphics[height=4.5cm,width=7cm]{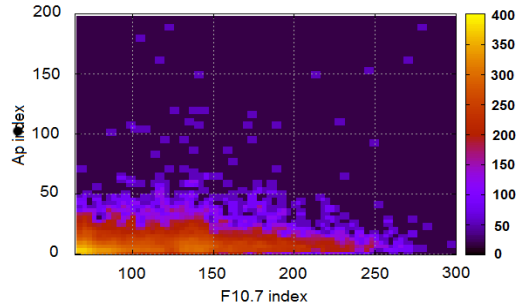}
\caption{Number of occurrences of solar events as a function of F10.7 \& Ap indices. Numbers are colour coded.
While the number is high for lower energetic events, those with very high values of solar parameters are sporadic and isolated}
\label{fig3}
 \end{center}
\end{figure}

The result of the statistical analysis of daily distribution of the solar flux (F10.7) 
and geomagnetic Ap index is shown in Fig. 2(a-d). 
The Fig. 2a is drawn for 1995-2009, the period between two minima. Figure 2b
is drawn for 1999-2013, the period between two maxima. They are almost identical. Figure 2c
is drawn for the period around the last solar maximum 1999-2003. Figure 2d is drawn for the period
around the solar minimum 2004-2008. The red points refer to the number of occurrences of a 
given F10.7 index and the blue points refer to the number of occurrences of a given Geomagnetic Ap Index.
Clearly, satellites and space probes would be most vulnerable to drags when both 
of these indices are simultaneously high. 
Major space weather events are accompanied by simultaneous
occurrences of VSA ($F10.7\geq200$) and SSC ($Ap\geq100$). The events including July 2000 solar 
event \cite{NASN04}, April 2001 solar event \cite{NASA05} and October/November 2003 
Halloween storm \cite{NASA06} during (and around) solar maximum are of this type. 
In Fig. 3, we draw the frequency of occurrences
in colors for any given pair of Ap index and solar F10.7 index. Clearly, number of 
events having both parameters high are rarer. Three possibilities 
could increase adverse space weather conditions and `magnetospheric' impacts.
They are: (i) High rate of CMEs (8-15/day) with average speed of $\geq 600km/s$, 
(ii) low/average rate of CMEs (3-7/day) with high/very high speed of about 1000-2250km/s, 
and (iii) high rate of CMEs (8-15/day with high/very high speed, which is the most 
damaging of all. On 14th July 2000, there were $5$ CMEs recorded including a 
halo CME with speed up to $1674$km/s, $7$ were recorded on 2nd April 2001 with 
speed up to $2505$km/s and about $24$ were recorded between 27th October and 
2nd November 2003, with several of them having a speed between $1000$ and $2598$km/s. 
Severe geomagnetic storms, both minor and major, also occurred during the solar minimum 
when `persisted' MSA ($F10.7\geq100$) occurred along with SSC ($Ap\geq100$). Nine severe 
geomagnetic storm conditions were recorded during 2004 to 2007; three in 2004, three in 2005 
and one in 2006. Analysis of the available data showed that 109 and 103 CMEs were respectively 
recorded in July and November 2004 during which the events occurred. Before the 
severe storms of 25th and 27th July, a Halo CME with a speed up to 1333km/s occurred 
on 25th July. Three severe storm conditions were recorded between 8th and 10th November, 
preceded by three halo CMEs with a speed between 1759 and 3387km/s. Out of the three severe 
storms recorded in 2005, more than seven Halo CMEs preceded the events with a speed 
between 1194 and 2250km/s. The only severe storm condition recorded in 2006 (on 15th Dec.)
was preceded by two Halo CMEs with speeds up to $1774$km/s.\\

\begin{figure}
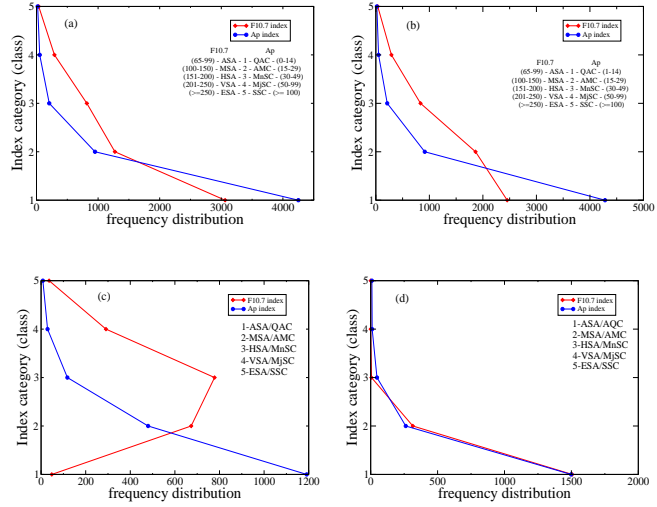

 \begin{center}
\includegraphics[height=3cm,width=4cm]{figure4a} \hspace{0.2cm}
\includegraphics[height=3cm,width=4cm]{figure4b} \\
\vspace{0.6cm}\includegraphics[height=3cm,width=4cm]{figure4c}
\hspace{0.2cm}\includegraphics[height=3cm,width=4cm]{figure4d}
 \end{center}
\caption{Frequency of occurrence of classified/grouped data (F10.7 \& Ap index) (a) min-to-min: 1995-2009 (b) max-to-max: 1999-2013 (c) around last solar maximum: 1999-2003 and (d) last solar minimum: 2004-2008}
\label{fig4}
\end{figure}

In Figs. 4(a-d), we plot the frequency distribution of F10.7 Index and Ap Index in four 
time slots we considered in Figs. 2(a-d).  In Y-axis, we plotted the categories defined in table \ref{tbl1}.
 Note that generally these two indices are correlated, i.e., both the indices
have a similar type of variations, though, during a minimum, the correlation is very tight.
Only exception is the statistics during the solar maximum - here the number of occurrences 
of events having very high and very low F10.7 indices is very low indeed, and the most probable
event being of category 3 (HSA). The Ap index, however, does to follow this pattern and 
distribution is monotonic.

\section{Computation of Orbital Decay of Satellites}

\begin{figure}
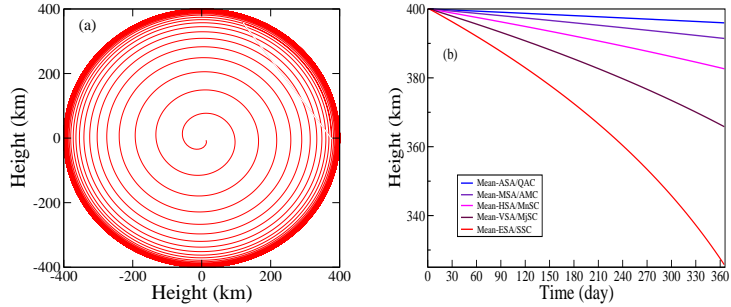

 \begin{center}
\includegraphics[height=4cm,width=4.5cm]{figure5a} \hspace{0.2cm}
\includegraphics[height=4cm,width=4.5cm]{figure5b}
 \end{center}
\caption{(a) Pictorial representation of decay motion of model satellite from 400km altitude, 
(b) orbital decay under the influence of mean value of classified/grouped space parameters (F10.7 and Ap index)}
\label{fig5}
\end{figure}

Figure 5(a-b) shows how a model satellite orbit decays with time from an initial altitude of 
$400$km. The average decay in one year under an average condition of each of the five categories of 
the solar events (as marked in the inset) was computed. The results are shown in 
Figure 5b. The reduction in height (decay) is about $4$, $9$, $18$, $35$ and $75$km respectively
for a condition of average, moderate, high, very high and extreme solar activity 
respectively. In Fig. 6(a-d), we present the result of two exercises.
In Fig. 6a, we present the solar flux (F10.7) and Ap index variation with time in the 1999-2003 period.
In Fig. 6b, we  present the the number of CMEs/Day and the linear speeds of the CMEs. 
In Fig. 6c, we show the results of our computation of the atmospheric drag on the satellite using
these observed quantities. Five curves show the results computed using the conditions in five successive `one year' interval (indicated) around the last solar maximum. The boxes in Fig. 6c, drawn around the regions of rapid decay in the orbits correspond to the major flare events presented in Figs. 6(a-b) by vertical boxes. 
In comparison, if we had chosen the data from a period of solar minimum, the results of 
decay in the period 2004-2008 would have been as shown in Fig. 6d. It is clear that 
in the period of solar minimum, the orbital decay per year is about half of what 
we obtain using parameters of the period of solar maximum.\\

\begin{figure}
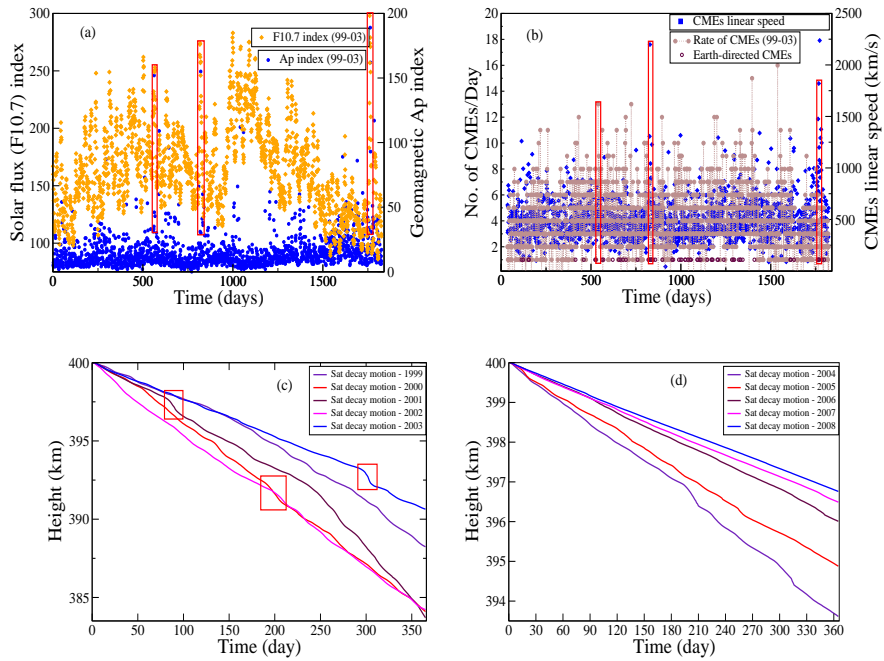

 \begin{center}
\includegraphics[height=4cm,width=5.5cm]{figure6a} \hspace{0.2cm}
\includegraphics[height=4cm,width=5.5cm]{figure6b} \\
\vspace{0.6cm}\includegraphics[height=4cm,width=5cm]{figure6c}
\hspace{0.35cm}\includegraphics[height=4cm,width=5cm]{figure6d}
 \end{center}
\caption{5-year daily distribution of (a) F10.7 and Ap index and (b) daily CME rate and mean linear speed, 
around last solar maximum with highlight of data around report dates of major solar events, 
(c) Satellite's orbital decay in one year interval during 1999-2003 
(d) Satellite's orbital decay in one year interval during last solar minimum, 2004-2008}
\label{fig6}
\end{figure}

\begin{figure}
 \begin{center}
\includegraphics[height=5cm,width=5.8cm]{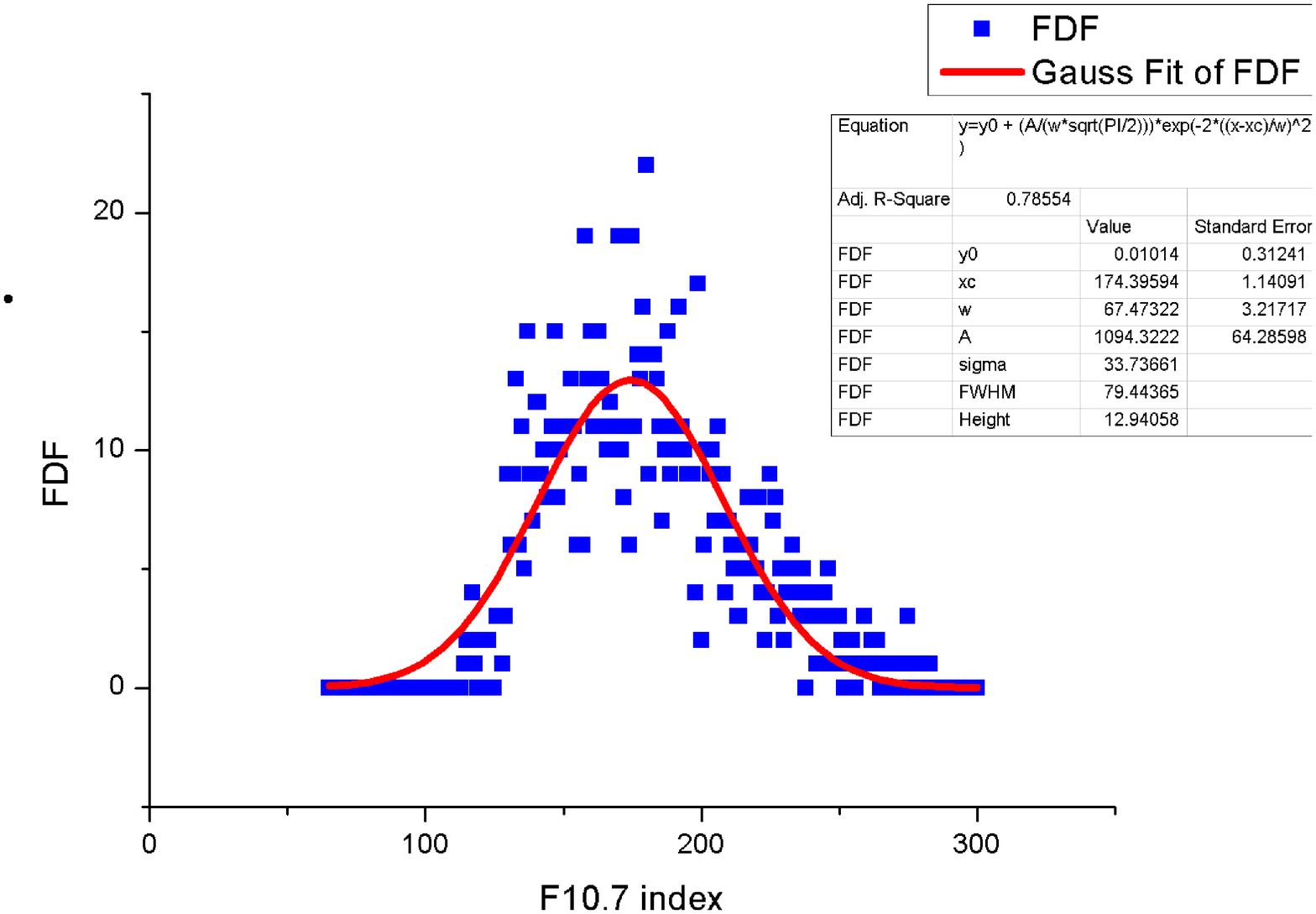}
\hspace{0.2cm} 
\includegraphics[height=5cm,width=5.8cm]{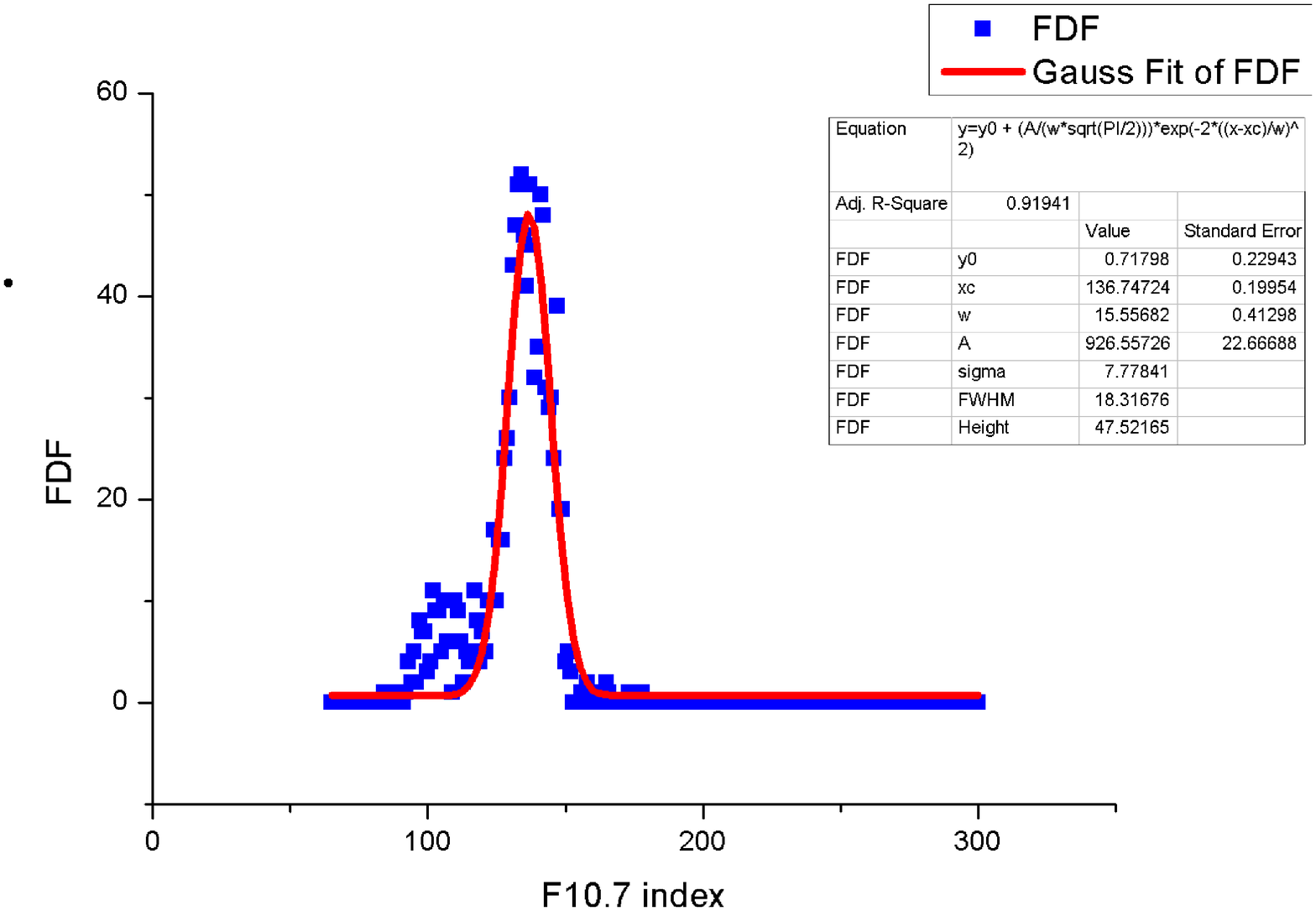} \\
\vspace{0.6cm}\includegraphics[height=4cm,width=5.5cm]{figure7c} \hspace{0.4cm}
\includegraphics[height=4cm,width=5.5cm]{figure7d} \\
\vspace{0.3cm}
 \end{center}
\caption{Frequency distribution with Gaussian fit of F10.7 index during 
(a) last solar maximum, 2000-2002 and (b) around next predicted maximum, 
2012-2014 (c) corresponding satellite's orbital decay motion during 2000-2002 and 
(d) 2012-2014} 
\label{fig7}
\end{figure}

It is interesting to carry out an exercise to study the dependency 
of the fate of a satellite orbit on the strength of the solar maximum. 
For the sake of concreteness, we compare two cases: (a) three year period during the last maximum, i.e., 2000-2002
and (b) three year period during the (predicted) upcoming maximum, i.e., 2012-2014.
Figures 7(a-b) show the frequency distributions with Gaussian fits of solar flux 
index for 3-year observed data during the last solar maximum (2000-2002) and emerging (observed + predicted) solar maximum (2012-2014) respectively. Values of 
F10.7 around 170 were the more frequent during last maximum, and during the next 
maximum about 136. The respective mean values of (F10.7, Ap) pairs are about 
(180,14) and (132,13) respectively (we assumed Ap ~13 during the next peak). 
The results of the corresponding orbital decays under the influence of these   
parameters are shown in Figs. 7(c-d). It is clear that the expected (F10.7, Ap) 
being less severe during the next peak, the effects of drag will be milder 
than the last solar maximum. The mean solar index for the two maxima fall 
in HSA and MSA categories respectively. Their average decay (in orbit) 
is about 80km and 35km, corresponding to $20\%$  
and $8.75\%$ drop in altitude within a period of three years. The predicted decay, however, 
excludes contribution from any possible major events during the period. In fact, we estimate 
that if we assume solar events similar to those of 2000, 2001 and 2003  in the
next solar maximum, then in addition to the average effect we computed above (Fig. 7d), 
we would have at least $2.5\%$ drag effect in excess of the average effect we presented 
above. Thus we predict that the net plasma drag effect on the orbital decay would be 
about $11.25\%$ during the next solar maximum.

\section{Conclusion}

In the present paper, we studied the effects of adverse space weather condition
induced by solar events, especially the solar flares and coronal mass ejections
on the orbital decay of low earth satellite orbits. We clearly show that the the result strongly 
depends on the phase of the solar cycle. First we studied the statistics of the 
F10.7 flux index and the Ap index since 1999 and systematically tracked the 
evolution of an hypothetical satellite in different phases of the solar cycle. 
Not surprisingly, we found that the effects in the last solar maximum 
was very severe, while expected effects occurring
from the ongoing cycle would be almost half as severe as the previous one.
We showed that a major CME event can cause sufficient heating and expansion of the atmosphere  
so that the orbital radius may go down by a few km in a single event. 

Our analysis of statistics of solar events indicates that all types of 
magnetic conditions, such as, QAC, AMC, MnSC, MjSC and SSC, in the 
magnetosphere can occur during any stage of the cycle. While QAC category events may occur
at any phase of the cycle, the geomagnetic activity could be significantly high 
during  a solar maximum such as the past cycle (1999-2003).
We find that during the last solar maximum, QAC was lower by about $20\%$, 
while AMC, MnSC and MjSC were higher by about $46\%$, 
$60\%$ and $62\%$ respectively as compared to the solar minimum stage (2004-2008). Naturally, 
we find that a satellite deployed during a solar minimum has higher chance of survival.

Based on estimations of the decay, we find that three types of space weather conditions
are potentially harmful to any space probe. (i) High rate of CMEs ($8-15$/day) 
with an average speed of $\geq 600$km/s; (ii) low/average rate of CMEs ($3-7$/day) 
with a high to very high speed of about $1000-2250$km/s; or, (iii) high rate of CMEs 
($8-15$/day) with high/very high speed. We have also observed a `double-peak' feature
during the period of solar maximum of a solar cycle. This means that the satellites are 
likely to be harshly affected for a longer period of time. One of our interesting findings 
is that for a typical satellite launched at a height of 400km, the plasma drag could 
cause up to $11.25\%$ decay of its orbit during the upcoming solar maximum,
with as much as about $2.5\%$ contribution coming from major CMEs.

\section*{Acknowledgment}
VUN acknowledges TWAS/ICTP, Trieste, Italy and the S.N. Bose National Centre For Basics Sciences
for the post graduate fellowship during which this work was done.

%\begin{itemize}
%\item Parenthetical: \verb|\cite{WB96}| produces \cite{WB96}.
%\item Textual: \verb|\citet{WB96}| produces \citet{WB96}.
%\item An affix and part of a reference:
%   \verb|\cite[e.g.][Ch. 2]{WB96}|
%   produces \cite[e.g.][Ch. 2]{WB96}.
%\end{itemize}

%%%%%%%%%%%%%%%%%%%%%%%%%%%%%%%%%%%%%%%%%%%%%%%%%%%%%%%%%%%%%%%%%%%%%%%%%%%%%
%% Appendices
% \appendix

\end{document}